\documentclass[twocolumn,nobalancelastpage,prl]{revtex4-1}
\usepackage[utf8]{inputenc}
\usepackage [english] {babel} 
\usepackage{amsmath,amssymb,latexsym}
\usepackage{slashed}
\usepackage{graphicx}
\usepackage{epstopdf}
\usepackage{mathtools}

\usepackage{xcolor}
\usepackage{hyperref}

\hypersetup{
  colorlinks=false
  urlbordercolor=gray,
  citebordercolor=gray,
  linkbordercolor=gray,
  pdfborderstyle={/S/U/W 1}
}

\DeclareGraphicsExtensions{.pdf}

\def \sec{\begin{section}}
\def \esec{\end{section}}

\renewcommand{\tilde}{\widetilde}

\def \al {\alpha}

\def \om {\omega}

\def \ep {\epsilon}
\def \th {\theta}

\def \Oc {\mathcal{O}}

\def \Fc {\mathcal{F}}

\def \Lc {\mathcal{L}}

\def \Cc {\mathcal{C}}

\def \altt {\tilde{\alpha}}

\def \Jtt {\tilde{J}}

\def \pr {\partial}
\def \ra {\rightarrow}

\def \beq { \begin{equation}}
\def \eeq {\end{equation}}

\DeclareMathOperator*{\Sch}{Sch}
\DeclareMathOperator*{\sgn}{sgn}

\renewcommand\Im{\operatorname{Im}}

\def \l {\left(}
\def \r {\right)}

\def \bra {\langle}
\def \ket {\rangle}

\def \nono {\nonumber \\}

\begin{document}
\title{Coupled Sachdev--Ye--Kitaev models without Schwartzian dominance}
\author{Alexey Milekhin}
\affiliation{Department of Physics, University of California at Santa Barbara, Santa Barbara, CA 93106,
U.S.A.}
\email{milekhin@ucsb.edu}
\begin{abstract}
We argue that in certain class of coupled Sachdev--Ye--Kitaev(SYK) models 
the low energy physics at large $N$ is governed by a
non-local action rather than the Schwartzian action. We present a partial analytic and extensive numerical
evidence for this. We find that these models are maximally chaotic and have the same residual entropy as Majorana SYK.
However, thermodynamic quantities, such as heat capacity and diffusion constant are different.
\end{abstract}

\maketitle
\section{Introduction}
Recently
Sachdev--Ye--Kitaev(SYK) model 
\cite{SachdevYe, parcollet1998, GPS, KitaevTalks,Gross:2016kjj}
and tensor models 
\cite{Gurau:2009tw,Witten:2016iux,Klebanov:2016xxf,Klebanov:2017nlk} 
have been the focus of much theoretical research. Also there have been several proposal of possible 
experimental implementation
\cite{Chew_2017,Danshita_2017,Chen_2018,Can_2019b,Wei_2021}.
The most prominent features of this family of models are non-Fermi liquid behavior,
maximal \cite{Maldacena:2015waa} Lyapunov exponent, non-zero residual entropy and
an approximate time reparametrization symmetry at large $N$ and low energies.
Maldacena and Stanford \cite{ms} and Kitaev and Suh \cite{SuhFirstPaper} derived the following (Euclidean)
Schwartzian action for reparametrizations in the original SYK model:
\begin{align}
\label{k:sch}
S_{Sch} = -\frac{N \alpha^S_{Sch}}{J} \int du \ \Sch\l \tau[u], u \r, \\
\Sch \l \tau[u],u \r =  \frac{\tau'''}{\tau'} - \frac{3}{2} \l \frac{\tau''}{\tau'} \r^2 \nonumber
\end{align}
In all known SYK-like modes, the Schwartzian has been identified as the dominant
\footnote{Here we talk about strict large $N$ limit. In certain tensor models subleading $1/N$ corrections
are different from SYK and are not captured by the Schwartzian.} low-energy action and
its physics has been explored in a variety of situations
\cite{MQ,Fu:2016vas,cotler2017black,gu2017spread,Yoon:2017nig,GurAri2018Does,
wormhole_form,syk_bath, bath_old, pengfei, KitaevRecent, su2021page}.
\textit{In this Letter we present a coupled SYK model which is dominated by
the following non-local action instead of the Schwartzian action:}
\beq
\label{k:s_nonloc}
S_{nonloc} = -\frac{N \alpha^S_{2h}}{J^{2h-2}} \int du_1 du_2  \l \frac{\tau'(u_1) \tau'(u_2)}{(\tau(u_1)-\tau(u_2))^2} \r^{h}
\eeq
It was suggested by Maldacena, Stanford and Yang \cite{cft_breaking} 
that this action may appear when a theory contains a local irrelevant operator with the
dimension $h$ within the interval $1<h<3/2$. 
Holographically this corresponds to a light matter field in $AdS_2$ having a source term at the boundary. 
In our model $h$ is tunable and can be
anywhere between $1$ and $2$. 

We would like to stress that the Schwartzian term is still present in our model.
The key point is that at large $N$ it gives a subleading in
$T/J$ contribution, where $T$ is the temperature.

Our approach is semi-analytical.
We consider strict large $N$ limit and obtain various predictions
of the non-local action. Then we check them against numerical solutions
of exact large $N$ equations. Also we demonstrate that the Lyapunov exponent is still maximal
and study the transport in chain models. 
In the accompanying longer paper \cite{soon} we provide more details on
numerics and theoretical computations, explore
other aspects of these models and study $1/N$ corrections.

\section{Microscopic formulation}
\label{sec:model}
The model consists of two independent Majorana SYK models($\Lc_{kin} + \Lc_{SYK}$),
a marginal interaction($\Lc_{int}$) between them, and an innocuously-looking kinetic term twist($\Lc_\xi$): 
\begin{align}
\label{eq:L_T}
\Lc_T = \Lc_{kin} + \Lc_{SYK} + \Lc_{int} + \Lc_{\xi},
\end{align}
where
\begin{align}
\Lc_{kin} = \frac{1}{2} \sum_i \l  \psi^1_i \pr_u \psi^1_i +  \psi^2_i \pr_u 
\psi^2_i \r
\end{align}
\begin{align}
 \Lc_{SYK} = \frac{1}{4!}\sum_{ijkl} \Bigg( J^1_{ijkl}  \psi^1_i \psi^1_j \psi^1_k \psi^1_l +  
J^2_{ijkl} \psi^2_i \psi^2_j \psi^2_k \psi^2_l \Bigg)
\end{align}
\begin{align}
\Lc_{int} = \frac{3}{2} \alpha \sum_{ijkl}
C_{ij;kl} \psi^1_i \psi^1_j \psi^2_k \psi^2_l 
\label{eq:Lint}
\end{align}
\beq
\label{eq:xi}
\Lc_{\xi} = -\frac{\xi}{2} \sum_i \l  \psi^1_i \pr_u \psi^1_i -  \psi^2_i \pr_u 
\psi^2_i \r
\eeq
There are $2N$ Majorana fermions $\psi^a_i,\ i=1,\dots,N,\ a=1,2$.
This action with $\xi=0$ has been studied in the literature before \cite{Gu2017Local,Altland:2019lne, bath_old} and
it was argued that it is dominated by the Schwartzian. It is important that tensors $J^{1,2},C$ are different.
When they are the same the ground state is actually gapped and the model is prone to $\mathbb{Z}_2$ symmetry breaking
\cite{Kim:2019upg}. Low energy
physics is not described by the conformal solution in this case. We will give a brief analytical argument
below why this does not happen in our model. Also we cross-checked this with exact diagonalization \cite{soon} at
finite $N$. 
The non-local action emerges only when both $\alpha \neq 0, \xi \neq 0$.
Non-zero $\alpha$ renders the two SYK models coupled and also it controls
the dimension $h$ in the non-local action. Specifically, we need $|\alpha|>1$ to make the non-local
action dominant.
Non-zero $\xi$ explicitly breaks $\psi^1_i \leftrightarrow \psi^2_i$ $\mathbb{Z}_2$ symmetry
and the needed irrelevant operator appears in the conformal perturbation theory(more on it below).
Tensors $J^1, J^2$, are standard SYK disorders(totally antisymmetric 
with i.i.d. Gaussian components).
Tensor $C_{ij;kl}$ has a Gaussian distribution too, but it has a separate skew-symmetry in $ij$ and $kl$ indices:
\beq
C_{ij;kl} = -C_{ji;kl} = -C_{ij;lk}
\eeq
With the variances:
\beq
\bra  \l J^a_{ijkl} \r^2 \ket = \frac{3! J^2}{N^3},\  a=1,2;\ \bra \l  C_{ij;kl} \r^2 \ket = \frac{J^2}{6 N^3}
\eeq
Euclidean Schwinger--Dyson(SD) equations read 
\begin{align}
\label{sd:eucl}
(1-\xi) \pr_u G_{11} - J^2(G_{11}^3 + 3 \alpha^2 G_{11} G_{22}^2) * G_{11} = \delta(u)  \nonumber \\
(1+\xi) \pr_u G_{22} - J^2(G_{22}^3 + 3 \alpha^2 G_{22} G_{11}^2) * G_{22} = \delta(u)
\end{align}
with $*$ denoting time convolution. $G_{11/22}$ are the time-ordered Green functions:
\beq
G_{aa}(u) = \bra T \psi^a_i(u) \psi^a_i(0) \ket,\ a=1,2
\eeq
In principle, $\xi$ can be reabsorbed into
$J^1,J^2$ variances. Mixed correlators $\bra \psi^1_i \psi^2_i \ket$ do not appear up to $1/N$ order.
These correlators serve as order parameters for the gapped  $\mathbb{Z}_2$-symmetry broken phase. This suggests that
this breaking does not happen in our model and the physics can be described by the following conformal solution.
At low energies we can neglect the kinetic term and $\xi$ disappears. 
Assuming $G_{11}=G_{22}$, one obtains the familiar SYK conformal
solution 
\begin{align}
\label{eq:Gconf}
G_{11,22} = G_{conf} = \frac{b \sgn(u)}{(1+3 \alpha^2)^{1/4}} 
 \frac{\pi}{\sqrt{ J \beta \sin\l \frac{\pi |u|}{\beta} \r }}
\end{align}
with $b = 1/(4 \pi)^{1/4}$. The full solution can be obtained by numerically
solving SD equations. Our numerical approach is a simple iteration procedure commonly used in SYK literature \cite{ms}.
We cross-checked our numerical solution against 
the results of exact diagonalization at finite $N$ \cite{soon}.

It is important that the equations are coupled, so there is only one reparametrization mode.
Time reparametrizations act on Green functions by
\begin{align}
\label{eq:reparam}
G_{aa}(u_1,u_2) \ra  \l \tau(u_1)' \tau(u_2)' \r^{1/4} & G_{aa}(\tau(u_1),\tau(u_2)), \nonumber  \\
& a=1,2 
\end{align}
This equation implies that $\psi$ have conformal dimension $1/4$.
It is easy to compute the dimension $h$  of the following bilinear operator
\cite{Kim:2019upg}:
\beq
\label{eq:theoperator}
\Oc_{2,0} = \sum_{i} \l \psi^1_i \pr_u \psi^1_i - \psi^2_i \pr_u \psi^2_i \r,
\eeq
it is given by the smallest solution of
\beq
\label{eq:dimension}
\frac{1-\alpha^2}{1+3 \alpha^2} g_A(h) = 1, \quad g_A(h) = -\frac{3}{2} \frac{\tan \l \pi(h-1/2)/2 \r}{h-1/2}
\eeq
For $|\alpha|>1$, the dimension $h$ is in the range: $1<h<3/2$.
This bilinear operator is exactly the $\Lc_\xi$ term in the Lagrangian. 
\section{A perturbative argument}
\label{sec:pert}
\begin{figure}[h]
\includegraphics[scale=0.4]{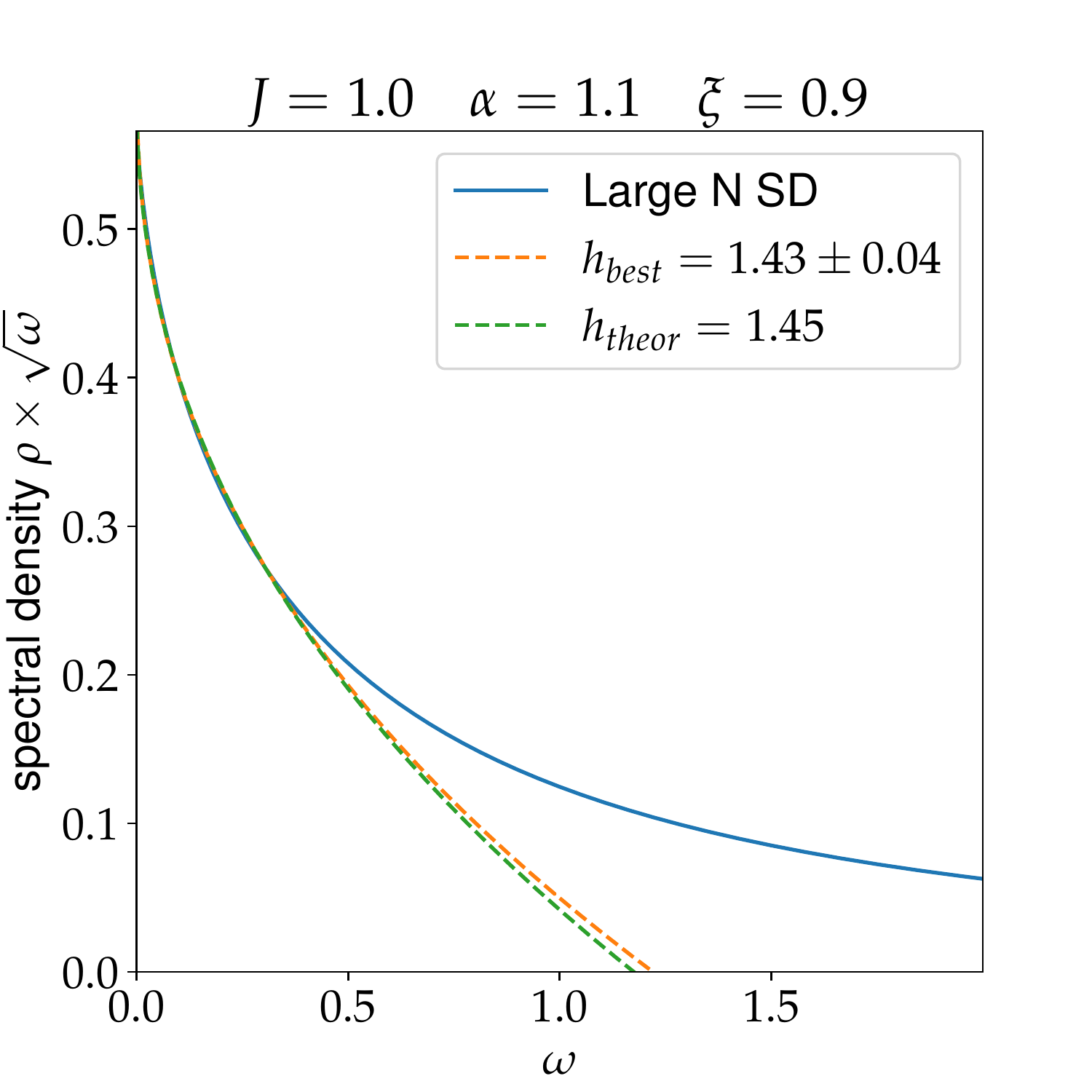}
\caption{Results for  $G_{11}$. The fit was 
performed with $b_1+b_2 \omega^{h-1}$. 
Uncertainty in $h_{best}$ results from changing the fitting interval.
Theoretical $h$ is obtained from eq.
(\ref{eq:dimension}). 
}
\label{fig:alpha_b1}
\end{figure}

As in the standard SYK, we can go beyond the conformal sector by perturbing \cite{SuhFirstPaper, Tikhanovskaya:2020elb} the
exact conformal Lagrangian
\beq
\Lc_{conf} = \Lc_{SYK} + \Lc_{int}
\eeq
by a set of irrelevant operators
\beq
\label{eq:irrels}
\Lc_{T} = \Lc_{conf} + \sum_h \alpha_{h} \Oc_h
\eeq
Unfortunately, there is no \textit{ab initio} way to compute $\alpha_h$.
The set of these irrelevant operators is constrained by the symmetries of the model.
For $\xi=0$ there is $1 \leftrightarrow 2$ symmetry at large $N$ and the operator 
(\ref{eq:theoperator}) does not appear. This is why we need the $\Lc_\xi$ term
in the UV Lagrangian. However, the corresponding IR parameter $\alpha_h$ 
does not have to be linear in $\xi$ \cite{soon}.
Now we can see the origin of the non-local action (\ref{k:s_nonloc}).
In perturbation theory, one can derive it by taking the 2-point
function of $\Oc_h$ and dressing it with reparametrizations \cite{cft_breaking}. Higher-order terms are suppressed by
$1/N$. This agrees with holographic expectations of weakly interacting fields in the bulk in the large $N$ limit.
Another consequence of this formalism is the leading non-conformal correction
to the 2-point function:
\beq
\label{eq:Gleading}
\delta G_{11}(u) = \alpha_h \int du' \ \bra \Oc_h(u') \psi^1_i(u) \psi^1_i(0) \ket \propto \frac{1}{(J u)^{h-1/2}}
\eeq
The last equality is valid for $|u| \ll \beta$. We checked this prediction by
numerically computing the spectral density at zero temperature and real frequency:
\beq
\rho_{11/22}(\omega) = \Im G_{R,11/22}(\omega),
\eeq
where $G_R$ is retarded Green function.
Conformal behavior (\ref{eq:Gconf}) results in $\rho \sim 1/\sqrt{\om}$ for $\om \ll J$, whereas the non-conformal
correction (\ref{eq:Gleading}) gives $\rho \sim \om^{h-3/2}$. Therefore, we expect the following behavior at small $\om$:
\beq
\sqrt{\om} \rho_{11/22}  = b_1 \pm b_2 \om^{h-1}
\eeq
Constant $b_1$ can be extracted from the conformal solution, but $b_2$ is related to $\alpha_h$ and 
has to be extracted from the numerics.
The result is presented in Figure \ref{fig:alpha_b1}. We see a good agreement with the numerical results.
%\end{widetext}

\section{Thermodynamics}
\begin{figure}[h!]
\includegraphics[scale=0.70]{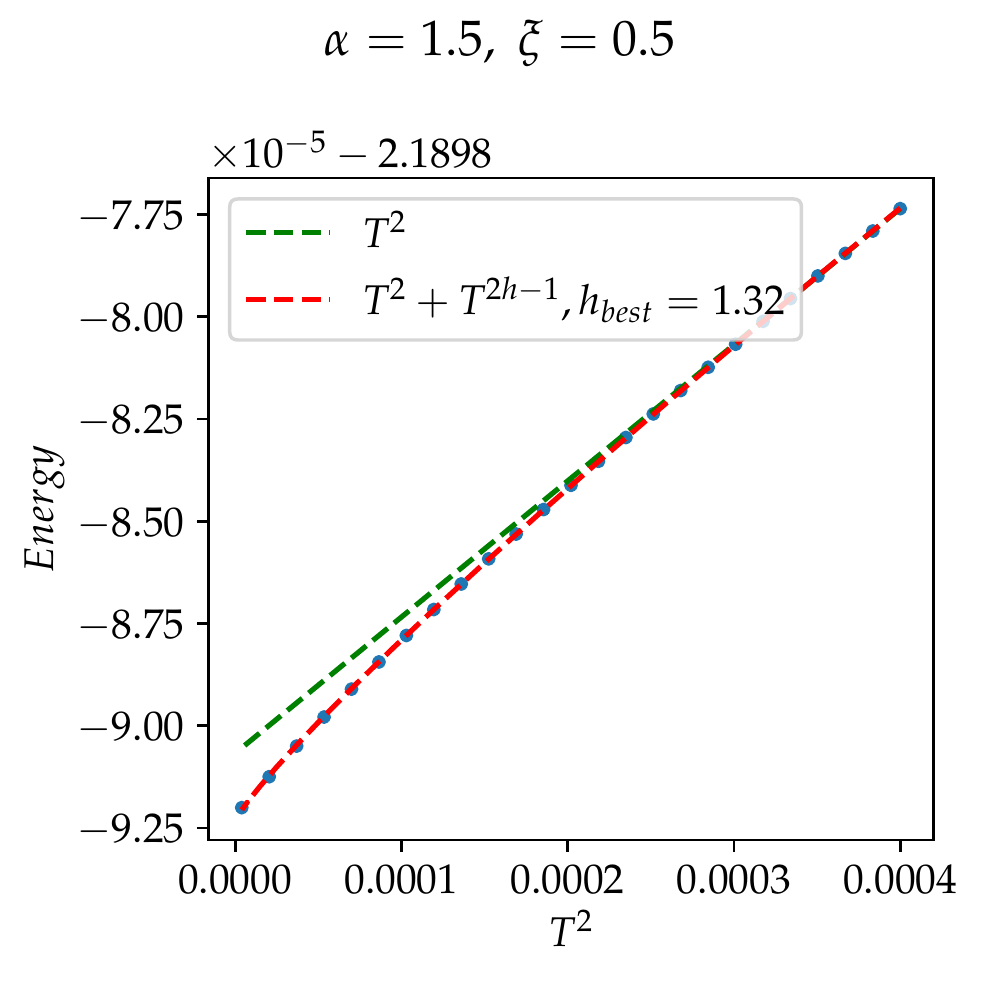}
\caption{Energy $E/N$ vs $T^2$ for $J=2\pi$. Blue points are numerical data. 
    There is a clear deviation from Schwartzian $T^2$ prediction.
    For $\alpha=1.5$,
    $h_{theor}=1.31$. Altering temperature range($\beta=50,\dots,500$), 
 the number of (uniform) discretization points($2^{23}-2^{26}$)
    and removing $c_{Sch} T^2$ term results in
    $h_{best}=1.32 \pm 0.02$. The fit was performed using eq. (\ref{eq:E_T}) with unknown $E_0, c_{2h}, c_{Sch}$.}
\label{fig:e_nonloc}
\end{figure}

Due to overall $N$ factor, we can treat the non-local action (\ref{k:s_nonloc}) classically
as long as $T \gg J/N^{1/(2h-2)}$. Its contribution
to free energy is equal to the classical action evaluated on the thermal solution $\tau(u) = \tan(\pi u/\beta)$.
This computation produces a divergence \cite{cft_breaking}:
\begin{align}
\label{eq:pred}
-\frac{\Delta F_{nonloc}}{N} 
%= \frac{\alpha_{2h}^S}{J^{2h-2}} \beta^2 
%\int_0^1 d\tilde{u}\ \l \frac{\pi}{\beta \sin^2 \l \pi \tilde{u} \r} \r^{2h} = \nonumber \\
= \# +   T^{2h-1} \frac{\alpha_{2h}^S \pi^{2h-1/2}}{J^{2h-2}} \frac{\Gamma \l 1/2-h \r}{\Gamma \l 1 - h \r }
\end{align}
where we assumed a fixed cut-off $u \sim 1/J$ in Euclidean time. The divergent term  $\#$
is temperature-independent and shifts the 
ground state energy, and so we drop it. Adding the Schwartzian contribution, we get the following answer for 
thermodynamic energy:
\beq
\label{eq:E_T}
E/N =E_0 + c_{2h} T^{2h-1} +  \frac{2\pi^2 \alpha^S_{Sch}}{J} T^2 + \dots
\eeq
where
\beq
c_{2h} = (2h-2) \frac{\alpha_{2h}^S \pi^{2h-1/2}}{J^{2h-2}} \frac{\Gamma \l 1/2-h \r}{\Gamma \l 1 - h \r }
\label{eq:FS}
\eeq
For $1<h<3/2$ the non-local piece dominates over the Schwartzian at low temperatures.
It is straightforward to extract the thermodynamic energy from numerically computed $G_{11}$ and $G_{22}$.
The comparison between the prediction (\ref{eq:E_T}) and the numerics
is shown in Figure \ref{fig:e_nonloc}, which demonstrates good agreement.

\begin{figure}[h!]
\includegraphics[scale=0.6]{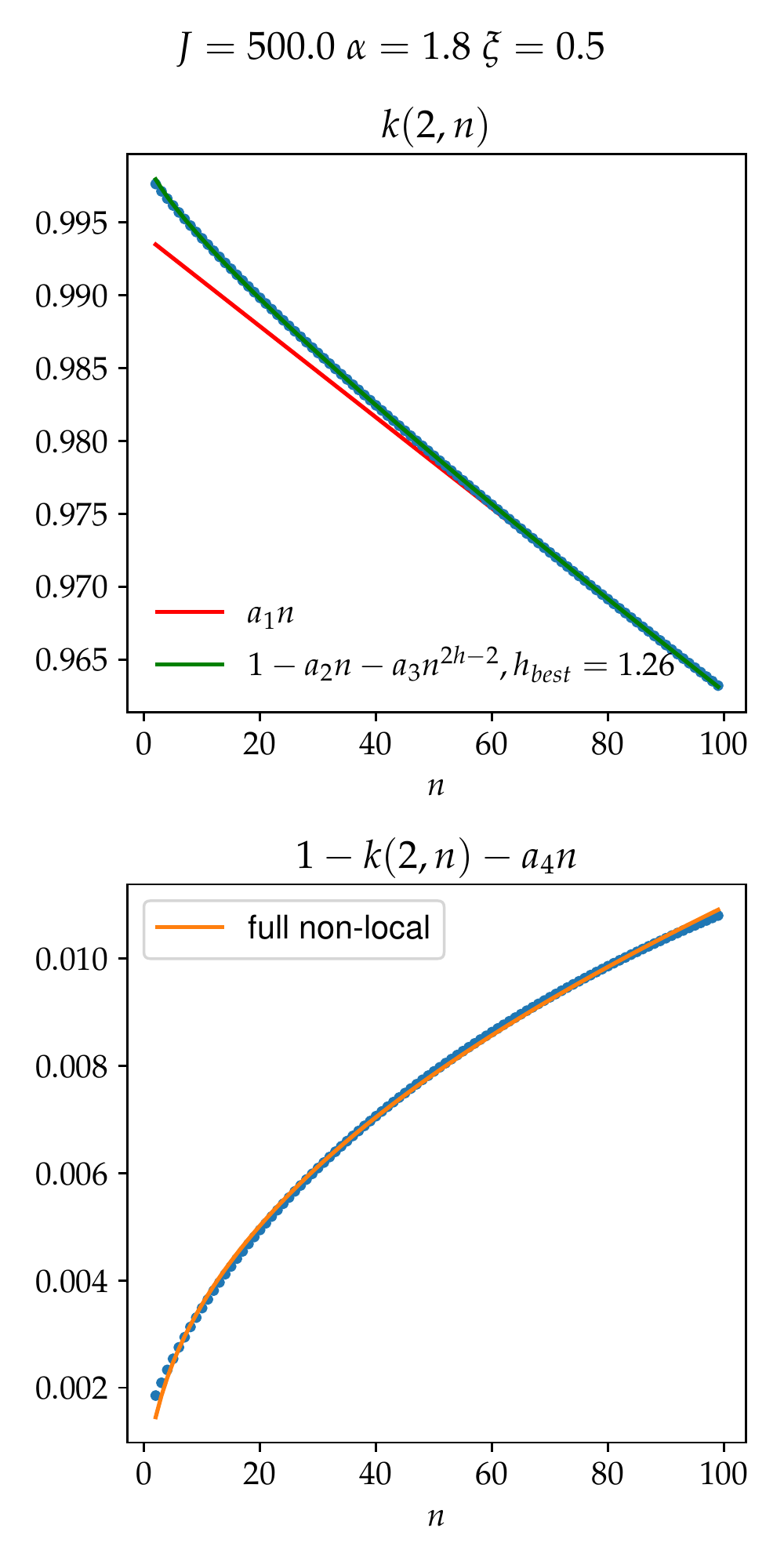}
    \caption{Results for $k(2,n)$. Blue points are numerically obtained eigenvalues of $K_n$ in eq. (\ref{eq:2Kernel_n}).
 Top: since $g_h(n)$ is actually numerically
 close to $n^{2h+1}$, we have performed the fit with $1-a_2 n - a_3 n^{2h-2}$ at large $n$ to extract $h_{best}$.
    For $\alpha=1.8$, $h_{theor}=1.24$, whereas $h_{best}=1.28 \pm 0.04$
Bottom: fit with eq. (\ref{xi:k}) with $\alpha^K_{Sch,2h}$ unknown: $a_4=\alpha^K_{Sch}/\beta J$.}
\label{fig:kns15}
\end{figure}
Conformal solution (\ref{eq:Gconf}) has the same form as in the original SYK. Zero temperature residual
entropy can be extracted from the conformal solution \cite{parcollet1998,KitaevRecent}.
Hence the residual entropy should be twice Majorana SYK residual entropy. Our numerical results support this claim.

\section{Kernel and 4-point function}
\label{sec:kernel}
As in the original SYK, connected 4-point function $\Fc$ is given by a sum of ladder diagrams:
$\Fc \propto 1/(1-K)$. "Flavor" indices $1,2$ turn the kernel $K$ into $2 \times 2$ matrix. It is the most convenient
to represent it by its action on vector $v=(v_1,v_2)$:
\begin{align}
    \frac{K v}{3J^2} =  \begin{pmatrix}
 G_{11} * \l \l G_{11}^2 + \alpha^2 G_{22}^2 \r v_{1} + \
 2 \alpha^2 G_{11} G_{22} v_{2} \r * G_{11}  \\
 G_{22} * \l \l G_{22}^2 + \alpha^2 G_{11}^2 \r v_{2} + \
 2 \alpha^2 G_{11} G_{22} v_{1} \r * G_{22}  \\
    \end{pmatrix}
\label{eq:2Kernel}
\end{align}
Using the translation symmetry, it is possible to separate the Matsubara frequency $n$  and re-write the kernel as a 
function of two times only \cite{KitaevRecent}:
\beq
\label{eq:2Kernel_n}
K_{n,ab}(u,u') = \int_0^{\beta} ds K_{ab}\l s  + \frac{u}{2}, s - \frac{u}{2}; \frac{u'}{2}, -
 \frac{u'}{2} \r e^{-2 \pi i n s/\beta}
\eeq
Because of the reparametrizations, this kernel has eigenvalue $1$ in the conformal approximation for any 
integer $n$ except $0,\pm 1$.
The eigenvalue shift $1-k(2,n)$ can be read off from the reparametrization action:
\beq
\label{xi:k}
1-k(2,n) = \frac{\alpha^K_{2h}}{(\beta J)^{2h-2}} 
\frac{g_h(n)}{|n|(n^2-1)} + \frac{\alpha^K_{Sch} |n|}{\beta J}
\eeq
\beq
\label{eq:gh}
g_h(n) = n^2 \l \frac{\Gamma(n+h)}{\Gamma(1+n-h)} -
\frac{\Gamma(h-1)}{\Gamma(-h)} \r 
\eeq
where the first term in $1-k(2,n)$ comes from the non-local action, 
and the second linear in $|n|$ term comes from the Schwartzian 
\footnote{The author thanks D.~Stanford and Z.~Yang for useful discussions about this
computation and the Lyapunov exponent computation.}. The relation between $\alpha^K_{2h}$ and $
\alpha^S_{2h}$ is
\beq
\alpha^K_{2h} = - (2 \pi)^{2h+1} \frac{\pi (h-1)^2 }{\cos(\pi h) \Gamma(2 h)}
\frac{4}{\pi^4 b^4} \alpha^S_{2h}
\eeq
Using the exact numerical solutions $G_{11,22}$ and fixing $\beta = 2\pi$,
 we build the kernel (\ref{eq:2Kernel_n}) for each $n=2,\dots,100$
and found numerically the eigenvalue $k(2,n)$ closest to $1$.
For this we used a uniform 2D grid in $(u,u')$ plane with $60,000^2-140,000^2$ points. The comparison between the prediction
(\ref{xi:k}) and the numerics is shown in Figure \ref{fig:kns15}. 

Performing the same procedure for different $J$, we verified $J^{2h-2}$ dependence in eq. (\ref{xi:k}) \cite{soon}.
The eigenvalue shift is sensitive to exact form of reparametrization action. We take the high levels of agreement with
the numerics as the main evidence for the non-local action dominance.

\section{Chaos exponent}
\label{sec:physics}
Let us demonstrate
that the Lyapunov exponent is maximal in the out-of-time ordered correlation function
(OTOC).
We are interested in the connected 4-point function 
\beq
\Fc =  \bra \psi^1_i(u_1) \psi^1_i(u_2) \psi^1_j(u_3) \psi^1_j(u_4) \ket_{conn}
\eeq
in the OTOC region:
\beq
u_1 = -\beta/4 +  i t, u_3 = 0,\ u_2 = \beta/4 + i t, u_4 = \beta/2
\eeq
The leading (in $1/\beta J$) contribution comes from averaging the disconnected 4-point function over the reparametrizations.
Delegating details to the Supplementary Material, we write down the final answer for OTOC:
\beq
\label{eq:otoc_fin}
\Fc = -p_h \frac{(\beta J)^{2h-2}}{N \alpha^S_{2h} m_h} 
 \exp \l \frac{2 \pi t}{\beta} \r + [\text{non-increasing}]
\eeq
with coefficient
\beq
p_h = \frac{\pi^{3-2h} \Gamma(2-h)}{4\Gamma(1+h) ( \psi(1+h) - \psi(2-h) )} G_{conf}(\beta/2)^2
\eeq
and where $\psi(x)=\Gamma'(x)/\Gamma(x)$.
The Lyapunov exponent $2 \pi t/\beta$ is maximal. However, the prefactor $(\beta J)^{2h-2}$
is smaller compared to Schwartzian-dominated original SYK, where it is $(\beta J)^1$.

\section{Transport in chain models}
\label{sec:conduct}
\begin{figure}[t]
\includegraphics[scale=1.6]{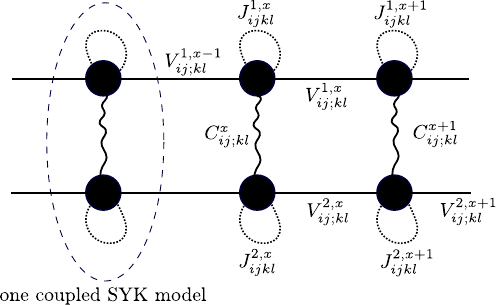}
\caption{Illustration of couplings in the chain model. }
\label{fig:chain}
\end{figure}

We can arrange the coupled model into an array and study transport properties.
In SYK chain models dominated by the Schwartzian \cite{Gu2017Local,Song_2017}, 
the diffusion constant is temperature-independent
and the heat conductivity is linear in the temperature. Here we find that the non-local action
renders the diffusion constant temperature-dependent, but leaves the heat conductivity proportional to the
temperature.
Specifically, we consider the following construction - Figure \ref{fig:chain}:
\beq
\mathcal{L}_{T, chain} = \sum_x \Lc_{T,x} + \Lc_{int,chain}
\eeq
where $\Lc_{T,x}$ is an independent copy of the coupled model (\ref{eq:L_T}).
The interaction between the sites should be carefully chosen in order to avoid a
possible spontaneous $\mathbb{Z}_2$ symmetry breaking. We study the following
interaction term between the sites:
\begin{align}
\label{eq:chain_int}
\Lc_{int, chain} = \frac{1}{2!^2} \sum_{x,ijkl} \Big( V^{1,x}_{ij;kl} 
 \psi^1_{i,x} \psi^1_{j,x} \psi^1_{k,x+1} \psi^1_{l,x+1} + \nonumber \\
V^{2,x}_{ij;kl} \psi^2_{i,x} \psi^2_{j,x} \psi^2_{k,x+1} \psi^2_{l,x+1} \Big)
\end{align}
Each $V^{1/2,x}_{ij;kl}$ i.i.d. Gaussian and is skew symmetric in $ij$ and $kl$:
\beq
V^{1/2,x}_{ij;kl} = - V^{1/2,x}_{ji;kl} = -V^{1/2,x}_{ij;lk},
\eeq
and has $x-$independent variance
\beq
\bra \l V^{1/2,x}_{ij;kl} \r^2 \ket = \frac{V^2}{N^3}.
\eeq
Assuming the two-point functions to be $x$-independent we arrive at single-site SD equations (\ref{sd:eucl}) with effective $\Jtt$ and $\altt$:
\beq
\Jtt^2 = J^2 + V^2,\ 
\altt = \al \frac{J^2}{J^2+V^2}
\eeq
It is important to keep in mind that the conformal dimension $h$ in the non-local action is now determined by this
renormalized $\altt$. In Supplementary Material we derive the following low-energy action for reparametrizations:
\begin{widetext}
\beq
S_{hydro} =  \frac{\pi^4 b^4 N}{4} \frac{\beta}{2 \pi} \int d\om dp \ \ep_{\om,p} \l \frac{\al^K_{2h}}{(\beta \Jtt)^{2h-2}} 
\frac{(2h-1) \Gamma(h)}{\Gamma(2-h)} \frac{\om^2 \beta^2}{4 \pi^2}  +
i p^2 \frac{\om \beta}{2 \pi} \frac{V^2}{3 \Jtt^2(1+3 \altt^2)} \r \ep_{-\om,-p}
\label{eq:hydro}
\eeq
\end{widetext}
The action is valid in the hydrodynamic regime $\om \ll T, p \ll 1$, where $p$ is the momentum conjugate to $x$. 
In this regime the reparametrization modes $\ep_{\om,p}$ are proportional to stress-energy tensor \cite{soon,Song_2017}.
We clearly see a diffusion pole in $\ep$ propagator with
the diffusion coefficient:
\beq
\label{eq:D}
D =  T^{3-2h} \frac{2 \pi \Gamma(2-h)}{3(2h-1) \Gamma(h)} \frac{V^2 \tilde{J}^{2h-2}}{\alpha^K_{2h} \tilde{J}^2(1+3\tilde{\alpha}^2)}
\eeq
This result differs a lot from SYK chains where the Schwartzian dominance 
results in temperature-independent diffusion constant.
Standard hydrodynamic expectation is that thermal conductivity is given by diffusion constant times the heat capacity.
This can be seen explicitly by the aforementioned mapping of $\ep_{\om,p}$ to stress-energy tensor.
In our case the heat capacity $c_v$ is proportional to $N T^{2h-2}$, eq. (\ref{eq:E_T}).
\textit{Therefore the thermal conductivity is linear in the temperature,
as in Schwartzian-dominated theories:} 
\beq
\kappa = c_v D \propto N T
\eeq

\section{Conclusion}
\label{sec:conclusion}
In this Letter we studied a coupled SYK model which shares a lot properties with the original SYK:
the same conformal solution, residual entropy and maximal chaos exponent.
However, in contrast to all known SYK-like models, the low energy physics at large $N$ is dominated 
by the non-local action for reparametrizations
(\ref{k:s_nonloc}) instead of the Schwartzian action. The reason behind this is the presence of the light conformal field
$\Oc_{2,0}$, eq. (\ref{eq:theoperator}),
with the dimension $h$, $1<h<3/2$. This conformal field is explicitly ``brought to life'' by the 
$\xi$ term in the Lagrangian (\ref{eq:L_T}). We would like to emphasize that this situation is not exotic:
we studied the simplest coupled SYK
and it is pretty easy to manipulate with conformal dimensions in the coupled SYK-like models 
\cite{Kim:2019upg,Klebanov:2020kck}.
Actually in our model for $|\alpha|<1$, the key operator $\Oc_{2,0}$, eq. (\ref{eq:theoperator}),
has the dimension $3/2<h<2$. This does not result in the dominance over the Schwartzian, but does result in
the leading non-conformal correction being different from SYK \cite{soon}.

The non-local action, compared to the Schwartzian, produces a different heat capacity, eq. (\ref{eq:E_T}),
OTOC prefactor, eq. (\ref{eq:otoc_fin}), and temperature-dependent diffusion constant in chain models, eq. (\ref{eq:D}).
Interestingly, the thermal conductivity is still linear in the temperature.
This result suggests that this linearity is a more general feature which perhaps can
be derived from the conformal solution. It would be very interesting to consider models with $U(1)$
symmetry and see if the electrical conductivity is linear in the temperature too as in strange metal.

Obviously, there are a lot of other open questions. 
There exists huge literature on SYK and the Schwartzian. Essentially, all the questions asked there could be asked
for the non-local action.
The most interesting of them is to study the strongly-coupled
regime of this model, when the temperature $T \lesssim J/N^{1/(2h-2)}$ and the non-local action
is not semiclassical. Up to non-perturbative $e^{-N}$ correction is this equivalent to quantizing 2D Jackiw--Teitelboim
gravity on a disk with light matter fields.

\section{Acknowledgment}
The author is forever indebted to I.~Klebanov, G.~Tarnopolsky and W.~Zhao for a lot of discussions and comments
throughout this project. It is pleasure to thank  A.~Gorsky, J.~Turiaci and especially D.~Stanford and 
Z.~Yang for comments, and F.~Popov for useful input and insightful suggestions on the paper.
I would like to thank C.~King for moral support and help with the manuscript.
This material is based upon work supported by the Air Force Office of Scientific Research under 
award number FA9550-19-1-0360. It was also supported in part by funds from the University of California. 
Use was made of computational facilities purchased with funds from the National
Science Foundation (CNS-1725797) and administered by the Center for Scientific
Computing (CSC). The CSC is supported by the California NanoSystems Institute
and the Materials Research Science and Engineering Center (MRSEC; NSF DMR
1720256) at UC Santa Barbara.

%\begin{thebibliography}{1}
%\end{thebibliography}
\bibliography{refs}

\begin{thebibliography}{39}
\bibitem [{Notea()}]{Notea}%
  \BibitemOpen
  \bibinfo {note} {Here we talk about strict large $N$ limit. In certain tensor
  models subleading $1/N$ corrections are different from SYK and are not
  captured by the Schwartzian.}\BibitemShut {Stop}%
\end{thebibliography}

\appendix
\section{Supplementary material}
\subsection{OTOC computation}
Averaging the product of two Green functions over the reparametrizations we get
\begin{align}
& \frac{\bra \psi^1_i(\th_1) \psi^1_i(\th_2) \psi^1_j(\th_3) \psi^1_j(\th_4) \ket_{conn}}{G_{conf}(x) G_{conf}(x')}
 = \frac{(\beta J)^{2h-2}}{N}  \frac{\pi^{2-2h}}{2 m_h \alpha_{2h}^S} \times \nono
& \times \sum_{|n| \ge 2} \frac{e^{i n (y'-y)}}{g_h(n)}
\left[ \frac{\sin \frac{n x}{2}}{\tan \frac{x}{2}} -
n \cos \frac{n x}{2}   \right]
\left[ \frac{\sin \frac{n x'}{2}}{\tan \frac{x'}{2}} -
n \cos \frac{n x'}{2}   \right]
\label{eq:4pt_raw}
\end{align}
$\theta_i$ are variables on the thermal circle, $\theta= 2\pi u/\beta$, and  $y,y',x,x'$ are
their combinations:
\beq
x= \th_1 -\th_2, \ x'=\th_3 - \th_4,\ y=\frac{\th_1+\th_2}{2},\ y'=\frac{\th_3+\th_4}{2}
\eeq
The computation simplifies a lot for points antipodal on the thermal circle. For them
\beq
\th_1 = -\frac{\pi}{2} - \th, \th_3 = 0,\ \theta_2 = \frac{\pi}{2}-\th, \th_4 = \pi
\eeq
Therefore we have the 4-point function (\ref{eq:4pt_raw}) proportional to
\beq
\sum_{|n| \ge 2} \frac{e^{i n\l \pi/2  + \th \r} n^2 \cos^2 \frac{\pi n}{2}}{g_h(n)}
=\oint_\Cc dn \ \frac{n^2}{e^{i \pi n}-1} \frac{e^{i n\l \pi/2  + \th \r}}{2 g_h(n)} 
\eeq
where the contour $\Cc$ encloses $\pm 2, \pm 4, \dots$. Pulling the contour to infinity
picks up zeros of $e^{i \pi n}-1$ and $g_h(n)$. 
We will be interested in the 
analytic continuation $\th \ra -2\pi i t/\beta$ to OTOC region. 
The only exponentially growing contribution comes from $n=1$. Computing the residue at $n=1$ yields
the expression (\ref{eq:otoc_fin}) in the main text.

\subsection{Hydrodynamic action in chain models}
To derive the action for reparametrization we need to study the kernel. Now there are two types of ladder diagrams:
on-site and the ones jumping between sites: $x \ra x \pm 1$. Taking a Fourier transform in $x$, we arrive at the
following result for the kernel:
\beq
K_{chain} = K_{ren} + K_p,
\eeq
where $K_{ren}$ is exactly on-site kernel (\ref{eq:2Kernel}), with renormalized $J,\alpha$.
$K_p$ is proportional to $\cos(p)-1$:
\begin{widetext}
\beq
K_p v  = 
2 V^2 (\cos(p)-1)\begin{pmatrix}
G_{11} * (G_{11}^2 v_{1}) * G_{11} & 0 \\
0 & G_{22} * (G_{22}^2 v_{2}) * G_{22} \\
\end{pmatrix}
\eeq
\end{widetext}
Separately, $K_{ren}$ and $K_p$ has an eigenvalue corresponding to reparametrizations.
One subtlety is that $K_p$ and $K_{ren}$ are not proportional as matrices, hence its hard to find the
eigenvalues of $K_{chain}$ precisely. 
We can ignore this issue in the long-wavelength limit $1-\cos(p) \approx p^2/2 \ll 1$, and just use the
conformal eigenvalue for $K_p$. The eigenvalue shift for $K_{ren}$ is controlled by reparametrizations, exactly
as in a single coupled model. Putting everything together, we have
\begin{widetext}
\beq
S =  \frac{\pi^4 b^4 N}{4}\sum_{n,p} \ep_{n,p} \l \frac{\al^K_{2h}}{(\beta \Jtt)^{2h-2}} g_h(n) + 
p^2 |n| (n^2-1) \frac{V^2}{3 \Jtt^2(1+3 \altt^2)} \r \ep_{-n,-p}
\eeq
\end{widetext}
Analytically continuing from the upper-half plane and keeping only $n^2$ term in $g_h(n)$ we get  the
hydrodynamic action (\ref{eq:hydro}).
%\printbibliography
\end{document}